\documentclass[12pt]{article}
\usepackage{latexsym}

\input epsf
\newcommand{\Tr}{\mbox{Tr\,}}
\newcommand{\beq}{\begin{equation}}
\newcommand{\eeq}[1]{\label{#1}\end{equation}}
\newcommand{\bea}{\begin{eqnarray}}
\newcommand{\eea}[1]{\label{#1}\end{eqnarray}}
\renewcommand{\Re}{\mathrm{Re}\,}
\renewcommand{\Im}{\mathrm{Im}\,}
\begin{document}
\hfill NYU-TH/00/09/09 hep-th/0009198

\vspace{20pt}

\begin{center}
{\Large \textbf{HOLOGRAPHIC DUALS OF 4D FIELD THEORIES~\footnotemark}}
\footnotetext{To appear in the proceedings of the conference 
{\it Strings, Duality and Geometry}, Montreal, March 2000.}
\end{center}

\vspace{6pt}

\begin{center}
\textsl{M. Porrati and A. Starinets} \vspace{20pt}

\textit{Department of Physics, NYU, 4 Washington Pl, New York NY 10012}

\end{center}

\vspace{12pt}

\begin{center}
\textbf{Abstract }
\end{center}

\vspace{4pt} {\small \noindent 
We discuss various aspect of the holographic correspondence between 
5-d gravity and 4-d field theory. First of all, we describe 
deformations of ${\cal N}=4$ Super Yang-Mills (SYM) theories
in terms of 5-d gauged supergravity. In particular, we describe
${\cal N}=0$ and ${\cal N}=1$ deformations of 
${\cal N}=4$ SYM to confining theories.
Secondly, we describe recent proposals for the holographic dual of 
the renormalization group and for 4-d central charges associated to it.
We conclude with a ``holographic'' proof of the Goldstone theorem.}

\vfill\eject 
\noindent
The ADS/CFT duality~\cite{m,gkp,w1} and its extension to non-conformal 
theories (see~\cite{agmoo} for a comprehensive review of the subject) 
has emerged over the last two years as a powerful tool
for understanding strongly coupled field theories. The best studied duality
is that between ${\cal N}=4$ SYM theory with gauge group $SU(N)$ and coupling 
constant $g_{YM}$, and type IIB superstrings on $AdS_5\times S_5$ in the limit
\beq
N\rightarrow \infty,\qquad \lambda=g_SN=\mbox{ constant}.
\eeq{1}
In this duality, when the 't Hooft coupling $g_SN=g_{YM}^2N$ is large, the
curvature of $AdS_5\times S_5$ is small ($\sim (g_SN)^{-1/4}$) so 
that $\alpha'$ corrections to type IIB supergravity, i.e. to the 
low-energy effective action of type IIB superstring, are small. In the large 
$N$ limit the string coupling $g_S$ is vanishingly small; therefore, 
string loop corrections are also small.
This ensures that semiclassical type IIB supergravity is a reliable 
approximation precisely when perturbative field theory fails.

Gauge-invariant operators of 4-d SYM are related to 10-d fields of type IIB
superstring on $AdS_5\times S_5$ (see~\cite{gkp,w1} for more details). Since
$S_5$ is compact one can expand the 10-d fields in 5-d KK states 
propagating on $AdS_5$. To identify a 5-d KK mode with a 
4-d operator they must transform identically under $SU(4)$. $SU(4)$ is both
the isometry group of $S_5$ (with spinors) and the R-symmetry group of 4-d SYM.
The conformal dimension $\Delta$ of a  4-d operator  
is determined by the mass of the corresponding 5-d KK mode.
For a scalar, it is the largest root of the equation~\cite{gkp}
\beq
\Delta(\Delta -4)= (ML)^2;\qquad L=(4\pi g_SN\alpha'^2)^{1/4}.
\eeq{2}

Since $L$ is the $AdS_5$ radius, a further simplification occurs in the 
$g_SN\rightarrow \infty$ limit. The mass of all excited string states is 
$O(\alpha'^{-1/2})$ so that their dimension $\Delta$
diverges in the limit and they decouple from the CFT.
Only the conformal dimension of states with mass $O(1/L)$ remains finite in 
the limit. These are precisely the KK modes of the states of 10-d 
type IIB {\em supergravity} (i.e. the states belonging to the 10-d 
graviton supermultiplet).   

This result fits nicely with superconformal field theory expectations, since 
(only) the KK states fit 
into short multiplets of the ${\cal N}=4$ superconformal 
algebra and consequently their conformal dimension is protected by a 
non-renormalization theorem. 

Among the scalar operators belonging to short multiplets, 42 are particularly 
interesting. They are associated to 5-d KK modes that survive a dimensional 
reduction to 5-d. In other words, 10-d type IIB supergravity on 
$AdS_5\times S_5$ can be consistently truncated to a 5-d supergravity on 
$AdS_5$ that contains only these 42 fields, together with their 
partners under 5-d ${\cal N}=8$ supersymmetry. No other truncation to a 
finite subset of KK modes exists.

These KK modes correspond to 4-d composite operators in the SYM 
theory~\cite{w1,ffz,gppz1}.

The ${\cal N}=4$ supermultiplet in 4-d contains one vector $A_\mu$, four
spin-1/2 fermions $\lambda^I$ and six scalars $\phi^A$, all in the adjoint of 
the gauge group ($SU(N)$ in our case).
$I$ labels the $\underline{4}$ and $A$ labels the $\underline{6}$ 
of the R-symmetry group $SU(4)$.
Under R-symmetry, the 42 operators decompose into a real 
$\underline{20}$, of conformal dimension 2, a complex $\underline{10}$, 
of conformal dimension 3, and a complex singlet of conformal dimension 4.
The dimension-2 operators are symmetric, traceless tensors of 
$SO(6)\sim SU(4)$: $\Tr \phi^{(A}\phi^{B)_T}$. The dimension-3 operators are
made of a fermion mass terms plus scalar trilinears, both symmetric tensors of
$SU(4)$: $\Tr \lambda^I \lambda^J + O(\phi^3)$. The dimension-4 operator is 
simply the ${\cal N}=4$ Lagrangian plus theta-term.

A deformation of ${\cal N}=4$ SYM by the 42 operators just discussed above can
be described using the dimensional reduction of 10-d type IIB supergravity
to 5-d gauged supergravity~\cite{grw}.  This theory has a 
complicated potential with several stationary points, 
besides the $SU(4)$-invariant one. The holographic correspondence suggests that
each (stable) stationary point of the potential describes a conformal 
field theory~\cite{gppz1,kpw}. A relevant deformation of ${\cal N}=4$ SYM 
generates a flow to another --possibly trivial-- local CFT. 
The holographic equivalent of this RG flow
is an appropriate solution of the equations of motion of 5-d supergravity.
Since we do not want to break Poincar\'e invariance, the ansatz for the
5-d metric is
\beq
ds^2= dy^2 + e^{2\phi(y)}dx^\mu dx_\mu,\qquad \mu=0,1,2,3.
\eeq{3}
The coordinate $y$ plays the role of RG scale~\cite{gppz1}, with larger $y$
corresponding to higher energy. The background
corresponding to a conformal field theory is an $AdS_5$ metric, with 
all 42 scalars $\lambda^a$ at a stationary point of the potential.

The only nonzero fields in our background are the metric and the scalars, so
that the relevant part of the 5-dimensional supergravity action is
\beq
L =\sqrt{-g}\left[
{R\over 4} + {1\over 2}\sum_a (\partial_y \lambda^a)^2 +V(\lambda^a)\right].
\eeq{mn9}
Here we have chosen for simplicity and without loss of generality 
canonical kinetic term for all scalars. 
Einstein's equations and the equations of motion of
the scalars following from eq.~(\ref{mn9}) are:
\beq
\partial_y^2\lambda^a + 4\partial_y\phi \  
\partial_y\lambda^a= {\partial V \over \partial \lambda^a},
\;\;\; 6(\partial_y\phi)^2=\sum_a (\partial_y\lambda^a)^2 -2V.
\eeq{m2}

Eqs.~(\ref{m2}) have solutions interpolating between two $AdS_5$ 
regions~\cite{gppz1}, as well as a universal runaway solution, 
independent of the detailed form of the potential~\cite{gppz2}.

In both cases for $y\rightarrow +\infty$ the solution asymptotes to the
$SU(4)$-invariant, ${\cal N}=4$ stationary point:
\beq
\lim_{y\rightarrow+\infty} \lambda^a(y)=0,\qquad 
\lim_{y\rightarrow+\infty}\phi(y)/y=1/L.
\eeq{4}

In the solutions of ref.~\cite{gppz1}, the metric is $AdS_5$ also in the limit
$y\rightarrow -\infty$ (with a different {\em larger} curvature). 
Those solutions are probably pathological because their $y=-\infty$ (IR) 
stationary points are non-supersymmetric and unstable~\cite{dz}. 

Ref.~\cite{gppz2} exhibited a different solution, singular in the IR. 
In that solution the scalars and the Einstein metric are
complicated and non-universal, {\em but their infrared behavior
is universal}. As shown in~\cite{gppz2}, whenever the metric and scalars become
singular at $y=a$, and whenever the scalar kinetic term is more singular than
the potential, one finds that the metric eq.~(\ref{3}) has the following
universal behavior:
\beq
ds^2= dy^2 + |y-a|^{1/2}dx^\mu dx_\mu.
\eeq{m4}
Eq.~(\ref{m4}) agrees with
the near-singularity form of the 5-d Einstein-frame metric found in
refs.~\cite{w2,min,poly}. As the example in~\cite{w2} shows, the singularity
is sometimes an artifact of the 5-d Einstein frame.

The singular metric described here is dual to a deformation of ${\cal N}=4$
SYM to a confining theory. Confinement can be proven by studying the
Wilson loop using the technique of~\cite{m2} as shown in~\cite{gppz2}. 

A deformation that preserves ${\cal N}=1$ supersymmetry is non generic, thus,
the 5-d metric that gives a holographic description of the deformation is 
not of the form given in eq.~(\ref{m4}). 
In terms of ${\cal N}=1$ superfields, ${\cal N}=4$ contains a vector
superfield $V$ and three chiral superfields, $\Phi^i$, $i=1,2,3$, that 
transform in the $\underline{3}$ of $SU(3)$. $SU(3)$ is the subgroup
of the ${\cal N}=4$ R-symmetry $SU(4)$ 
that commutes with the ${\cal N}=1$ supercharge. 
We can deform ${\cal N}=4$ SYM to pure ${\cal N}=1$ SYM by adding the 
${\cal N}=1$ supersymmetric F-term $m\int d^2\theta \Tr \Phi^i\Phi^i$ 
to the ${\cal N}=4$ Lagrangian.

The 5-d field corresponding to this deformation is uniquely 
identified~\cite{gppz3} 
by first  decomposing the $\underline{10}$ of $SU(4)$ 
under $SU(3)$:
\beq
\underline{10}\rightarrow \underline{1} + \underline{3} + \underline{6},
\eeq{5}
and by further decomposing the $\underline{6}$ of $SU(3)$ as
$\underline{1}+\underline{5}$ under $SO(3)\subset SU(3)$.

As shown in ref.~\cite{freed1}, a background of 5-d supergravity that 
preserves ${\cal N}=1$ supersymmetry exists if the scalar potential
$V$ can be written in terms of a superpotential $W$ as
\beq
V=\frac{1}{8}\sum_{a=1}^{n}\left|
\frac{\partial W}{\partial \lambda^a} \right|^2
- \frac{1}{3} \left|W \right|^2,
\eeq{6}
and if the fields satisfy the first-order equation
\bea
\dot\lambda^a&=&\frac{1}{2} \frac{\partial W}{\partial \lambda^a},\\
\dot\phi&=& - \frac{1}{3} W.
\eea{7}
In our case we set to zero all scalar fields except $m$ and the $SU(3)$
singlet in the decomposition~(\ref{5}), hereafter called $\sigma$. 
The SYM operator corresponding to that field is $\Tr \lambda^4\lambda^4$, i.e.
the ${\cal N}=1$ gaugino condensate. 
In terms of these fields, the superpotential is~\cite{gppz3}
\beq
W=\frac{3}{4} \left(\cosh{\frac{2m}{\sqrt{3}}} + \cosh{2\sigma}\right).
\eeq{8}
The supergravity e.o.m.~(\ref{7}) can be solved exactly~\cite{gppz3}:
\bea
\phi(y)&=&\frac{1}{2}\log[2\sinh(y-C_1)] + \frac{1}{6}\log[2\sinh(3y-C_2)],
\label{9a}\\
m(y)&=&\frac{\sqrt{3}}{2}\log\left[\frac{1+e^{-(y-C_1)}}
{1-e^{-(y-C_1)}}\right], \label{9b}\\
\sigma(y)&=&\frac{1}{2}\log\left[\frac{1+e^{-(3y-C_2)}}
{1-e^{-(3y-C_2)}}\right].
\eea{9c}
This solution is singular, but this singularity is acceptable, as long as
$C_1\geq C_2/3$~\cite{g}. Here and below we have rescaled the $AdS_5$
radius to $L=1$.

The asymptotic UV behavior of $m(y)$ and $\sigma(y)$ is
\beq
m(y)\sim \sqrt{3}e^{C_1}e^{-y},\qquad 
\sigma(y)\sim e^{C_2}e^{-3y},\qquad y\rightarrow\infty.
\eeq{10}
These equations show that $m$ is a true deformation of ${\cal N}=4$ SYM, 
with UV scaling dimensions 3, and that $\sigma$ is indeed the VEV
$\langle \Tr \lambda^4\lambda^4\rangle$.

The latter identification deserves an explanation, as it will be useful later.

A supergravity scalar $\lambda$ of mass $M$ behaves at large $y$ as
\beq
\lambda(y)= \lambda_0    e^{(\Delta-4) y} + C e^{-\Delta y}.
\eeq{11}
In the holographic interpretation, $\lambda_0$ is the source of a 
dimension-$\Delta$ operator, $O$, and the partition function
$Z[\lambda_0]=\langle \exp(-\int d^4x \lambda_0 O)\rangle $ is given by 
the supergravity action $S[\lambda]$ computed at the stationary point with 
boundary condition 
$\lim_{y\rightarrow\infty}\exp[(4-\Delta)y]\lambda(y)=\lambda_0$~\cite{w1}:
\beq
\langle e^{-\int d^4x \lambda_0 O}\rangle = \left. e^{-S[\lambda]}
\right|_{\rm stationary \; point}.
\eeq{12} 
Choosing for simplicity a canonical kinetic term for $\lambda$, and 
substituting eq.~(\ref{11}) into eq.~(\ref{12}) we immediately find: 
\beq
\langle O \rangle = 
\left. {\delta S \over \delta \lambda_0} \right|_{\lambda_0=0}=
-\int dy {\partial \over \partial y} \left[ e^{4\phi(y)} e^{(4-\Delta)y} 
{\partial \lambda \over \partial y}\right]=\Delta C.
\eeq{13}
In the case of $\sigma$, $\Delta=3$, so that its asymptotic behavior --
$\sigma \sim \exp(-3y)$-- is the
correct one to generate a gaugino condensate.

For further details we refer the interested reader to ref.~\cite{gppz3}, which 
also contains a detailed study of the Wilson loop in the 
background given in eqs.~(\ref{9a},\ref{9b},\ref{9c}). 

We move now to a 
brief discussion of the RG equations in the holographic framework.

Let us restore dimensions to $y$ and introduce again the $AdS_5$ radius $L$.
The supergravity action is divergent on an asymptotically $AdS$ background.
To regularize it, one can excise the asymptotic region $y>L\log (L\Lambda)$;
$\Lambda$ is clearly a UV cutoff. When all fields are independent of the 4-d
coordinates, the supergravity action depends on the 
coordinate $y$ only through the scale factor $\phi(y)$. We will find it useful
to define a new coordinate $\mu=\exp[\phi(y)]/L\leq \Lambda$.  In terms of this
new coordinate, a generic 5-d metric can be written as
\beq
ds^2=\omega(\mu)^2d\mu^2 + \mu^2 g_{\mu\nu}(\mu,x)dx^\mu dx^\nu.
\eeq{14}
Here $\omega(\mu)=\mu^{-1} dy/d\phi$ and the metric $g_{\mu\nu}(\mu,x)$
is asymptotically flat: $g_{\mu\nu}(\mu,x)\approx \eta_{\mu\nu}$ at 
large space-like $x$.
A plausible definition of the holographic renormalization group is as follow 
(see also~\cite{dbvv,bk,ps}).
Define ``bare'' fields, independent of $\mu$ as 
\beq
\lambda^a_B(x)=\lambda^a(\mu=\Lambda,x),\qquad 
g_{B\,\mu\nu}(x)=g_{\mu\nu}(\mu=\Lambda,x).
\eeq{15}
Define also $\mu$-dependent ``renormalized'' fields as the fields 
$\lambda^a(\mu,x)$, $g_{\mu\nu}(\mu,x)$ that solve the supergravity e.o.m. 
with boundary conditions $\lambda^a_B(x)$, $g_{B\,\mu\nu}(x)$~\footnotemark.
\footnotetext{To be precise, we also need appropriate boundary conditions 
at the IR boundary. In all concrete cases in the literature, it is not 
difficult to find them out explicitly.}    
Cleary, the ``bare'' supergravity action is independent of $\mu$; therefore,
\beq
0=\mu {d \over d\mu} S[\lambda^a_R, g_{B\,\mu\nu},\mu]=
\mu{\partial \over \partial \mu} S + \int d^4x \left[{d \lambda^a_R \over d\mu}
{\delta S \over \delta \lambda^a_R}(x) + {d g_{B\,\mu\nu} \over d\mu}
{\delta S \over \delta g_{B\,\mu\nu}}(x)\right].
\eeq{16}
This equation is not a tautology once one gives independent 
equations for $\dot{\lambda}^a_R\equiv d \lambda^a_R /d\mu$ and 
$\dot{g}_{B\,\mu\nu}\equiv d g_{B\,\mu\nu}/ d\mu$ (the beta functions).
The beta function equations are obtained by splitting the supergravity action 
$S$ into its UV and IR parts:
\beq
S=S_{UV}[\mu]  + S_{IR}[\mu] = \int d^4 x \int_\mu^\Lambda d\nu{\cal L}(\nu,x)
+ \int d^4 x \int_{\mu_0}^\mu d\nu {\cal L}(\nu,x).
\eeq{17}    
Here, $\mu_0$ stands for the physical IR cutoff, $\mu_0=0$ if the IR theory 
is conformal.
Finally, the RG equations are
\beq
\dot{\lambda}^a_R(\mu,x)=-{\delta S_{UV} \over \delta \lambda^a_R}(\mu,x),
\qquad
\dot{g}_{B\,\mu\nu}(\mu,x)=-{\delta S_{UV} \over \delta g_{B\,\mu\nu}}(\mu,x).
\eeq{18}
Eqs.~(\ref{16},\ref{18}) define a holographic renormalization scheme where 
the beta functions are $\beta^a\equiv \dot{\lambda}^a_R(\mu,x)$, 
$\beta_{\mu\nu}=\dot{g}_{B\,\mu\nu}(\mu,x)$. Notice that because of the 
definition of the renormalized metric, $\beta_{\mu\nu}$ vanishes on 
translationally-invariant backgrounds.  

Eqs.~(\ref{16},\ref{18}) also suggest a natural candidate central function 
$c(\mu)$.

Let us briefly recall the properties of a central function. 
\begin{enumerate}
\item $c(\mu)$ must decrease along an RG trajectory: 
$c_{IR}\equiv\lim_{\mu\rightarrow 0}c(\mu)\leq \lim_{\mu\rightarrow \infty}
c(\mu)\equiv c_{UV}$.
\item $\dot{c}(\mu)=0$ at the fixed points of the RG group (conformal points).
\item At the fixed points, $c$ is one of the central charges of the 
conformal algebra. In CFT described by supergravity duals, i.e. at large 't 
Hooft coupling, there exists only one central charge~\cite{hs}
\beq
\langle T_\mu^\mu (x) \rangle = c \left( -{1\over 8} R^{\mu\nu}R_{\mu\nu}
+{1\over 24} R^2 \right).
\eeq{19}
\end{enumerate}

A central function obeying all these properties was found in~\cite{gppz1} 
(see also~\cite{freed1}):
\beq
c[\mu]= \mbox{const}\, \left(\dot{\phi}\right)^{-3}.
\eeq{20}
It is monotonic because of the following equation~\cite{agpz}, that can be 
taken as the definition of the holographic scheme
\beq
\dot{c}= 2c \dot{\lambda}^a \dot{\lambda}^b G_{ab}.
\eeq{21}
This equation follows from the supergravity e.o.m.~(\ref{m2}); here $G_{ab}$ is
the scalar kinetic term, not necessarily canonical. Monotonicity of $c$ along a
generic RG trajectory can also be proven using the null energy 
condition~\cite{freed1}. 

Our definition of $c$ is unique only at the critical
points $\dot{c}=0$. Away from criticality, $c$ need not coincide with
central functions defined in other ways; indeed, it need not coincide with
other holographic definitions of $c$, as for instance that of 
ref.~\cite{dbvv}. 
This non-uniqueness, even within the holographic scheme, is due to
the ambiguity in the identification of $\phi$ as a function of the
scale $\mu$. The standard identification $\phi = \log(\mu/\mu_0)$ is unique
only at the critical points $\dot{c}=0$, because of the AdS/CFT
correspondence. Away from criticality, uniqueness is lost.

A function that reduces to $c$ at the RG fixed points can be defined in any
field theory by
computing the two point function of the stress-energy tensor using the
equation~\cite{cfl} 
\begin{equation}
\langle T_{\mu\nu} (x) T_{\rho\sigma}(0)\rangle = -{\frac{1}{48\pi^4}}
\Pi^{(2)}_{\mu\nu\rho\sigma} \left[{\frac{c_H(x)}{x^4}}\right] +
\pi_{\mu\nu}\pi_{\rho\sigma}\left[{\frac{f(x)}{x^4}}\right],
\label{duepunti}
\end{equation}
where $\pi_{\mu\nu}=\partial_\mu\partial_\nu -\eta_{\mu\nu}\partial^2 $, and 
$\Pi^{(2)}_{\mu\nu\rho\sigma}=2\pi_{\mu\nu}\pi_{\rho\sigma} -3
(\pi_{\mu\rho}\pi_{\nu\sigma}+\pi_{\mu\sigma}\pi_{\nu\rho})$. We call
$c_H$ the \textit{canonical} $c$-function.

In a generic field theory, $c_H(x)$ is not monotonic~\cite{cfl}. 
In theories admitting a supergravity dual it is, as we shall now see.
The 
holographic correspondence eq.~(\ref{12}) extends straightforwardly
to $T_{\mu\nu}$ once we find the source that couples to 
the stress-energy tensor. To do that, we expand the metric 
as $g_{\mu\nu}(y,x)= 
\exp[2\phi(y)]\eta_{\mu\nu} + \delta g_{\mu\nu}$. For $y\rightarrow +\infty$ 
we have $\delta g_{\mu\nu}(y,x)=
\exp(2y/L)h_{\mu\nu}(x) + O(1)$ so that the source is  
$h_{\mu\nu}$. Eq.~(\ref{12}) now reads 
\beq
\langle e^{-\int d^x T^{\mu\nu}h_{\mu\nu}}\rangle = e^{-S[g_{mn}]},
\eeq{12a}
where $g_{55}=1$, $g_{\mu 5}=0$, and $g_{\mu\nu}$ 
obeys the boundary condition
\beq
\lim_{y\rightarrow +\infty}
e^{-2y/L}g_{\mu\nu}(y,x)= \eta_{\mu\nu} + h_{\mu\nu}(x).
\eeq{12b}
To find the two-point function of $T_{\mu\nu}$ we
compute the supergravity action to quadratic order in $h$
\beq
\langle T_{\mu\nu} (x) T_{\rho\sigma}(0)\rangle=\left.{\delta^2 S\over 
\delta h^{\mu\nu}(x) \delta h^{\rho\sigma}(0)}\right|_{h=0}.
\eeq{22}

The on-shell supergravity action has the following form
\beq
S[h_{\mu\nu}]=\int d^4x dy e^{4\phi(y)}\delta g_{\mu\nu}(y,x) \Box^{-1}
\Pi^{(2)\,\mu\nu\rho\sigma} \delta g_{\rho\sigma} + ...,
\eeq{23}
Ellipsis denote terms proportional to the trace of the metric.
Because of eq.~(\ref{23}), the transverse-traceless part in 
eq.~(\ref{duepunti}) can be written as $\Box^{-2} \Pi^{(2)}_{\mu\nu\rho\sigma}
G(x)$, where $G(x)$ is the boundary-to-boundary Green function of a 5-d
minimally coupled massless scalar.

$G(x)$ can be computed as follows.
Consider a minimally-coupled massless scalar propagating in the background
eq.~(\ref{3}). It obeys the equation of motion
\beq
\left(\partial_y e^{4\phi}\partial_y +e^{2\phi}\Box \right)\psi(y,x)=0.
\eeq{24}
Its Fourier transform near the boundary is
\beq
\tilde{\psi}(y,k)=\left[1 + a_1e^{-2y/L} k^2 + a_2ye^{-4y/L} k^4 + e^{-4y/L}
\tilde{G}(k)+ o(e^{-4y/L})\right] \tilde{\psi}(k);
\eeq{25}
$\tilde{G}(k)$ is the Fourier transform of $G(x)$.

A detailed calculation of $G(x)$ and the canonical $c_H(x)$ that 
results from it, in a few 
cases where the computation can be done analytically,
has been reported elsewhere~\cite{agpz} (see also~\cite{ps2}). 
For the two flows examined in 
ref.~\cite{agpz} it was found that $c_{H\,IR}\equiv
\lim_{|x|\rightarrow 0}c_H(x) < c_{H\, UV}\equiv \lim_{|x|\rightarrow \infty}
c_H(x)$. 
As we mentioned above, this property is not generic in 4-d CFT; 
counterexamples were found in~\cite{cfl}. In theories with holographic 
supergravity duals, though, $c_{H\,IR}\leq c_{H\, UV}$. This inequality is
obvious when $c_{H\,IR}=0$, since positivity of the two-point function
eq.~(\ref{duepunti}) implies $c_H(x)\geq 0$~\cite{cfl}.

When $c_{H\, IR}>0$, the IR fixed point is a nontrivial CFT. In this case,
the inequality  $c_{H\,IR}\leq c_{H\, UV}$ holds because 
at the fixed point $c_H(x)$ and $c(y)$
coincide~\cite{hs}, because $c(y)$ is monotonic (eq.~(\ref{21})), 
and because the usual UV/IR connection~\cite{m} holds near the fixed points:
$|x|\rightarrow \lambda |x| \sim y \rightarrow y - L\log \lambda $ 
for $|y|\rightarrow \infty$. The last fact is more or less obvious; 
but we can also prove the inequality quite easily as follows~\cite{ps}.

Let us define the quantity 
\beq
\tilde{G}_y(k)\equiv e^{4\phi(y)} {\partial_y \tilde{\psi}(y,k) \over 
\tilde{\psi}(y,k)}.
\eeq{26}
It obeys $\lim_{y\rightarrow\infty}\tilde{G}_y(k)=\tilde{G}(k) + ak^2 + bk^2$,
where $a$ and $b$ are constants.
Because of eq.~(\ref{24}),  $\tilde{G}_y(k)$ satisfies the equation~\cite{ps}
\beq  
\partial_y \tilde{G}_y(k) = k^2 e^{2\phi(y)} - 
e^{-4\phi(y)} [\tilde{G}_y(k)]^2.
\eeq{27}
Expanding $\tilde{G}(k)$ near $k^2=0$ we find $\Im \tilde{G}(k) = O(k^4)$
and $\Re \tilde{G}(k) = O(k^2)$.  
Keeping only the lowest non-vanishing terms in $k^2$ 
in  eq.~(\ref{27}) and using the initial conditions given above we find
\beq
\Re \tilde{G}_y(k) = O(k^2) ,\qquad
\partial_y \Im \tilde{G}_y(k) =  -2 e^{-4\phi(y)}\Re\tilde{G}_y(k)\Im
\tilde{G}_y(k)= O(k^2)\Im \tilde{G}_y(k).
\eeq{28}
This equation implies that $\Im \tilde{G}_y(k)= \Im\tilde{G}_y(k) + O(k^6)$.
Since $\Im  \tilde{G}(k) = (4\pi)^{-1} c_{H\, IR}k^4 +O(k^6)$ and
$\lim_{y\rightarrow -\infty}\Im  \tilde{G}_y(k) = (4\pi)^{-1} c_{IR}k^4 +
O(k^6)$~\cite{gkp},
we obtain $c_{H\, IR}=c_{IR}\leq c_{UV}=c_{H\,UV}$. The last equality is 
obvious.

Finally, let us give a holographic formulation of the Goldstone theorem.
The key ingredient here is that global symmetries of the 4-d theory 
correspond to 5-d {\em gauge} symmetries of the supergravity dual~\cite{w1}.
The boundary value $A_\mu(x)$ of the 5-d gauge field is the source of
the Noether current associated to the symmetry. 
Let us call $B$ the expectation value of an operator $O$ in the presence of 
a 5-d gauge field:
\beq
B=B^0 +\int d^4k B^1_\mu(-k)\tilde{A}^\mu(k) + O(\tilde{A}_\mu^2).
\eeq{29}
Here $A_5=0$ by gauge choice. The relation between $B$ and 
the asymptotic form of its associated 5-d field is given by eq.~(\ref{13}).
Since $A_\mu(x)$ is the source of the 
conserved current $J_\mu(x)$, $B^1_\mu(-k)$ is the two-point function 
$\langle \tilde{J}_\mu(-k)O(0)\rangle$.

A pure-gauge field, $\tilde{A}_\mu=k_\mu \Lambda (k)$, can be set to zero 
with a 
gauge transformation {\it that acts as the 4-d symmetry on} $O$. We find then
another expression for $B$:
\beq
B=B^0 +\int d^4k \delta B(-k)\Lambda(k) + O(\Lambda^2),\qquad
\delta B(-k)\equiv \langle \delta_\Lambda O\rangle.
\eeq{30}
If $\lim_{k\rightarrow 0}\delta B(-k)\equiv \delta B$ is nonzero, 
eqs.~(\ref{29},\ref{30}) imply $\lim_{k\rightarrow 0}k^\mu B^1_\mu(-k)=
\delta B\neq 0$. By Lorentz invariance $B^1_\mu(-k)=k_\mu B^1(-k)$ and
$B^1(-k)=\delta B/k^2$ for $k\rightarrow 0$. This means that the two-point
function $\langle \tilde{J}_\mu(-k)O(0)\rangle$ has a massless pole, physical 
since $J_\mu$ and $O$ are both gauge invariant.
Notice that the only point where we used holography was in the identification
of the source $A_\mu$ with a 5-d gauge field.

In this note, we have surveyed various aspects of the holographic duality 
between strongly interacting 4-d field theories and 5-d supergravity, and we 
have found the holographic dual of several features of field theory. This
``dictionary'' allows for the study of strongly interacting theories by means 
of classical, weak-curvature (super)gravity.
    
\vskip .2in
\noindent
{\bf Acknowledgments}\vskip .1in
\noindent
This work is supported in part by NSF grants PHY-9722083 and PHY-0070787.

\end{document}